\documentclass[aps,pra,twocolumn,superscriptaddress, nofootinbib, longbibliography]{revtex4-1}
\pdfoutput=1
\usepackage[utf8]{inputenc}
\usepackage[english]{babel}
\usepackage[T1]{fontenc}
\usepackage{amsmath}
\usepackage{hyperref}
\usepackage{tikz}
\usepackage{lipsum}
\usepackage{float}
\usepackage{graphicx}
\usepackage{latexsym}
\usepackage{amssymb}
\usepackage{csquotes}
\usepackage{nicefrac}
\usepackage{amsfonts}
\usepackage{enumerate}
\usepackage{upgreek}
\usepackage{subfigure}
\usepackage{placeins}
\usepackage{bm}
\usepackage[version=3]{mhchem}
\usepackage{nicefrac}
\usepackage{bbold}
\usepackage{verbatim}
\usepackage{multirow}
\usepackage{color}
\usepackage{comment}
\usepackage{xcolor}
\usepackage{soul}
\usepackage{bibunits}

\usepackage{braket}

\renewcommand{\vec}[1]{\mathbf{#1}}

\begin{document}

\title{Magnetic correlations in the $SU(3)$ triangular-lattice $t$-$J$ model at finite doping}

\author{Annika Böhler}
\affiliation{Department of Physics and Arnold Sommerfeld Center for Theoretical Physics (ASC), Ludwig-Maximilians-Universit\"at M\"unchen, Theresienstr. 37, M\"unchen D-80333, Germany}
\affiliation{Munich Center for Quantum Science and Technology, Schellingstr. 4, Munich D-80799, Germany}

\author{Fabian Grusdt}
\affiliation{Department of Physics and Arnold Sommerfeld Center for Theoretical Physics (ASC), Ludwig-Maximilians-Universit\"at M\"unchen, Theresienstr. 37, M\"unchen D-80333, Germany}
\affiliation{Munich Center for Quantum Science and Technology, Schellingstr. 4, Munich D-80799, Germany}

\author{Annabelle Bohrdt}
\affiliation{Department of Physics and Arnold Sommerfeld Center for Theoretical Physics (ASC), Ludwig-Maximilians-Universit\"at M\"unchen, Theresienstr. 37, M\"unchen D-80333, Germany}
\affiliation{Munich Center for Quantum Science and Technology, Schellingstr. 4, Munich D-80799, Germany}

\date{\today}
\begin{abstract}
Ultracold alkaline-earth atoms and molecules now enable experimental realizations of SU(N)-symmetric Fermi-Hubbard models, yet theoretical understanding of these systems, particularly at finite doping remains limited. Here we investigate the strong-coupling limit of the $SU(3)$ symmetric Fermi-Hubbard model on the triangular lattice
across the full doping range. Using a three-flavor extension of Gutzwiller-projected hidden fermion determinant states (G-HFDS), a neural network based variational ansatz, we analyze two- and three-point spin-spin and spin-spin-hole correlations of the $SU(3)$ Cartan generators.  We further study the structure of a pair of doped holes for large periodic systems, and compare our results to the paradigmatic $SU(2)$ square lattice equivalent, finding strikingly similar magnetic correlations, non-s-wave pairing symmetry, and enhanced binding energies. Our results provide a foundation for future exploration of doped SU(N) Mott insulators, providing valuable insights for both theoretical developments and quantum simulation experiments.

\end{abstract}
\maketitle

\textbf{Introduction.} The Fermi-Hubbard (FH) model plays a central role in understanding the physics of doped Mott insulators (MI), and has been extensively studied both numerically~\cite{Arovas2022, LeBlanc2015, Jiang2022, Qin2020, Zheng2017} and experimentally~\cite{Bohrdt2021}. At weak coupling, these systems can often be approached using perturbative methods, offering a controlled understanding of ordering phenomena~\cite{Raghu2010, Keimer2015}. However, key features such as magnetism, pairing, and pseudogap physics arise from strongly entangled degrees of freedom in the strong coupling regime. The need for a universal approach to strongly coupled systems motivates extensions to higher $SU(N>2)$ spin symmetries, which offer a controlled framework for disentangling the specific features of $SU(2)$ physics from more general aspects of strong couplings. $SU(N)$ versions of the FH model, relevant for ultra-cold atoms~\cite{Cazalilla_2014, Ibarra-Garcia-Padilla_2025} and multi-orbital materials~\cite{Tokura2000, Li1998}, reveal a rich variety of emergent phenomena beyond the physics of the $SU(2)$ systems~\cite{Assaad2005,Honerkamp2004,IGP2014,Sotnikov2014}. 

Experimental advances have enabled the realization of $SU(N)$ symmetric FH models using ultracold alkaline-earth atoms
~\cite{Ibarra-Garcia-Padilla_2025}.
These experiments have observed Mott insulating states and magnetic correlations in $SU(N=3,4,6)$ FH models~\cite{Tusi_2022, Taie_2012, Hofrichter_2016, Ozawa_2018, Taie_2022, Buob_2024, mongkolkiattichai2025quantumgasmicroscopythreeflavor}. Recent experimental progress further allows the study of $SU(N)$ symmetric systems with ultracold molecules, enabling an even broader class of symmetries up to $SU(N=36)$, due to their rich internal structure~\cite{Mukherjee_2025}. While at unit filling $\langle \hat{n}_i\rangle=1$ and large $U$, $SU(N)$ Heisenberg models have been shown to reveal intricate magnetic orders~\cite{Toth2010, Nataf2014, Bauer2012, Harada2003, Nataf2016, Corboz2011}, finite doping studies of these models remain limited~\cite{Schloemer2024, Feng2023, IGP2023}.

In this work we employ a neural quantum state (NQS) ansatz~\cite{Carleo_2017} to study the strong coupling limit of the $SU(3)$ Fermi Hubbard model. Hereby, the wavefunction is parameterized as a neural network (NN), leveraging its universal function approximation capabilities~\cite{Hornik1991}. The ability to approximate the ground state wave function has been successfully applied to areas challenging for traditional methods, such as frustrated spin systems~\cite{Rende2024, chen2025convolutionaltransformerwavefunctions, roth2021group}, volume-law entangled states~\cite{Sharir2020, Deng2017, Levine2019} and more recently also fermionic and bosonic systems~\cite{lange2024architecturesapplicationsreviewneural, nomura2024quantummanybodysolverusing, Romero_2025, lange2024simulatingtwodimensionaltjmodel, Robledo_Moreno_2022, chen2025neuralnetworkaugmentedpfaffianwavefunctions, gu2025solvinghubbardmodelneural}.

We argue that the $SU(3)$ triangular lattice shares key physical features with the paradigmatic $SU(2)$ square lattice, such as the existence of a three-flavor Neel state at zero doping that breaks the $SU(3)$ symmetry. The elementary charge carriers of both models are described by magnetic polarons, and we compare the evolution of polaron correlations for the full doping range across both systems, providing microscopic explanations in terms of geometric strings~\cite{Chiu_2019, Grusdt_2019}. We further analyze pairing energies and symmetries of a pair of holes in the $SU(3)$ model and find enhanced pairing compared to the $SU(2)$ square lattice case. 
\\

\begin{figure}[t]
    \centering
    \includegraphics[width=0.48\textwidth, trim=0cm 0cm 0cm 0cm]{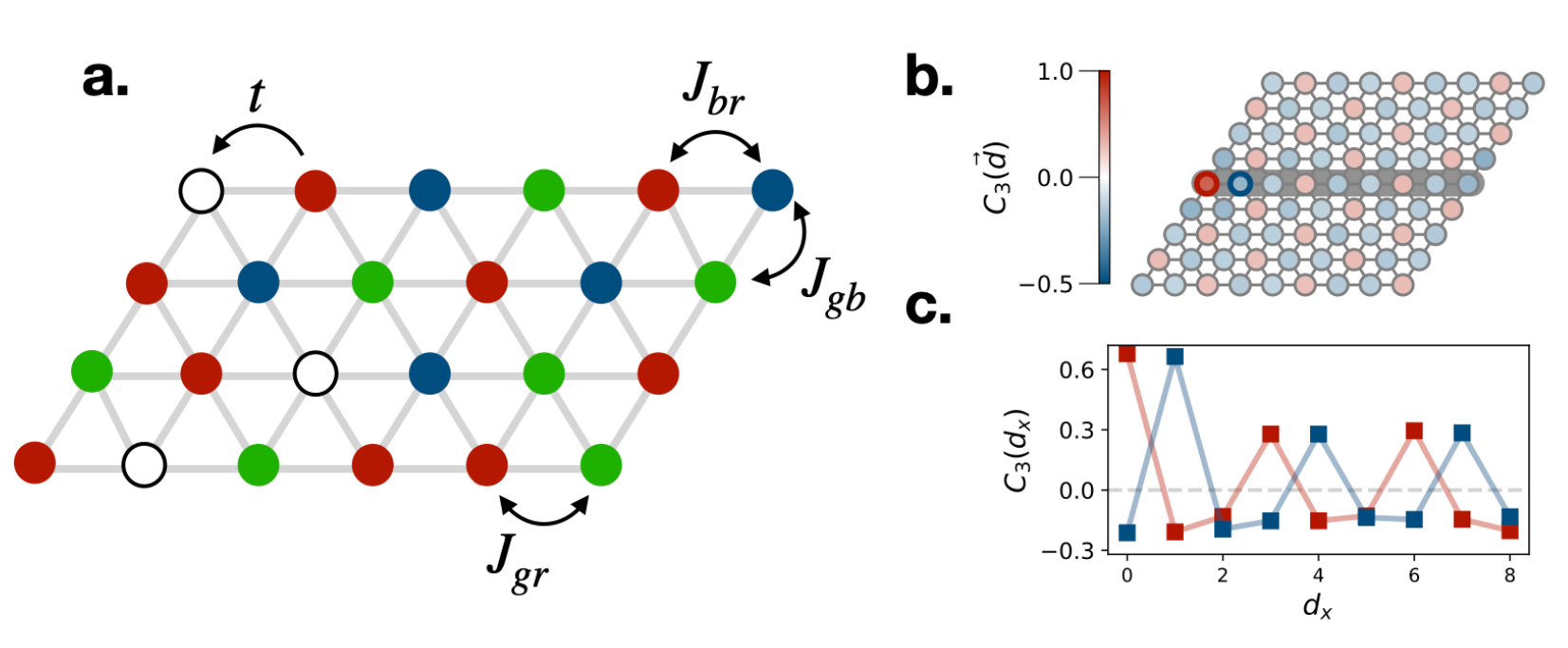}
    \caption{\textbf{a.} $SU(3)$ $t$-$J$ model, as given in Eq.~\eqref{eq:H_tJ}, with the $SU(3)$ symmetry requiring $J_{rg}=J_{br}=J_{gb}$. \textbf{b.}
    Spin-Spin correlation map $\langle \hat{\lambda}_{3, \mathbf{i}}\hat{\lambda}_{3, \mathbf{i+\vec{d}}}\rangle$ of the undoped system at unit filling on a $9\times 9$ torus. 
    \textbf{c.} Cut along the highlighted row in \textbf{b}, showing spin-spin correlations for different reference sites marked in \textbf{b}. The three-site periodicity for both reference sites confirms the three-sublattice order.
    }
    \label{fig:1}
\end{figure}

\textbf{Model and Neural Network Ansatz.}
We study the strong coupling limit of the $SU(3)$ symmetric FH model, the $SU(3)$ $t$-$J$ model
\begin{align}
\label{eq:H_tJ}
\begin{split}
    \hat{H}^{SU(3)}_{tJ}=&-t \sum_{\langle \mathbf{i},\mathbf{j} \rangle} \hat{\mathcal{P}} \sum_{\alpha} \left( \hat{c}_{\mathbf{i}\alpha}^\dagger \hat{c}_{\mathbf{j}\alpha}+h.c. \right)\hat{\mathcal{P}} \\ &+\frac{J}{2}\sum_{\langle \mathbf{i},\mathbf{j}\rangle}\hat{\mathcal{P}} \left( \sum_{\alpha,\beta}  \left( \hat{c}_{\mathbf{i}\alpha}^\dagger \hat{c}_{\mathbf{i}\beta} \hat{c}_{\mathbf{j}\beta}^\dagger \hat{c}_{\mathbf{j}\alpha}\right) - \hat{n}_\mathbf{i}\hat{n}_\mathbf{j}
\right) \hat{\mathcal{P}},
\end{split}
\end{align}

\noindent on the triangular lattice, where the sum over $\alpha$ runs over three spin flavors, here represented by red, green and blue, see Fig.~\ref{fig:1}a. The local particle number is given as $\hat{n}_\mathbf{i} = \sum_\alpha \hat{n}_{\mathbf{i} \alpha}$ and  $\langle \mathbf{i},\mathbf{j}  \rangle$ denotes nearest neighbors on the triangular lattice. $\hat{\mathcal{P}}=\prod_i\prod_{\langle \alpha, \beta \rangle}(1-\hat{n}_{i\alpha}\hat{n}_{i\beta})$ represents a Gutzwiller projection onto maximally singly occupied sites, where $\langle \alpha, \beta \rangle$ represents pairs of flavors.

At unit filling, the model maps to the $SU(3)$-symmetric Heisenberg model, whose ground state has been shown both numerically and analytically to exhibit three-sublattice (3-SL) order on both square and triangular lattices~\cite{Toth2010,Bauer2012}. Contrary to the $SU(2)$ triangular lattice the $SU(3)$ model does not exhibit geometric frustration. At zero doping, the ground state is an $SU(3)$ symmetry-breaking antiferromagnet, described by the three-flavor Néel state,
in close analogy to its counterpart on the $SU(2)$ square lattice
~\cite{Brauner_2010}. This is to be contrasted with the $SU(3)$ square lattice case, where a macroscopic ground state degeneracy prohibits the existence of a classical Néel state, making it numerically intractable without a priori inducing the correct symmetry breaking~\cite{Bird_2025, Schloemer2024}. In the low-doping limit of the AFM, the elementary charge carriers of both models can be described as magnetic polarons, consisting of doped charges dressed by a cloud of spin excitations. 


\begin{figure*}[t]
    \centering
    \includegraphics[width=0.99\textwidth]{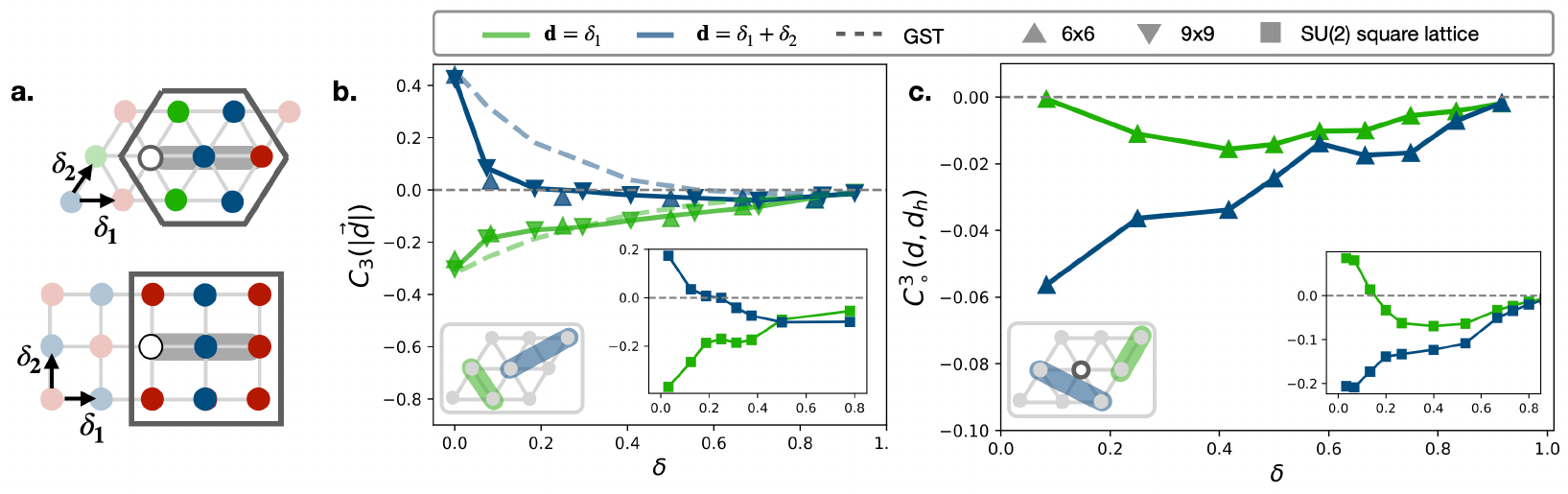}
    \caption{\textbf{a.} String patterns in the $SU(3)$ triangular and $SU(2)$ square lattice.
    \textbf{b.} Nearest and next-nearest neighbor spin-spin correlations $C_3(d)$ as a function of hole-doping $\delta=N_h/L^2$. We compare nearest-neighbor (\textit{nn}) sites on different sublattices and next-nearest-neighbor (\textit{nnn}) sites on the same sublattice. The inset shows a comparison with the same observable on the $SU(2)$ square lattice~\cite{lange2024simulatingtwodimensionaltjmodel}
    We compare magnetic correlations derived from the optimized hidden fermion determinant state (HFDS) with geometric string theory predictions (dashed, GST). The GST values are calculated from snapshots taken from the HFDS at half filling and modified as described in the main text. \textbf{c.} Connected nearest and next-nearest neighbor correlations $C^3_\circ(d,d_h)$ for $d_h=1$, as indicated in the cartoon inset. We average over all pairs of \textit{nn} and \textit{nnn} sites around the hole. The inset plot shows the analogous correlations for the $SU(2)$ square lattice case.}
    \label{fig:2}
\end{figure*}




In this study, we use of an NN ansatz with a hidden fermion construction~\cite{Robledo_Moreno_2022}.
We use a feed-forward neural network (FFNN) with a single hidden layer to represent the hidden fermion part of the ansatz, and impose an additional symmetrization over all possible spin permutations, i.e. $\psi(s) = \sum_{g\in S_3} \det M(gs)$, where $S_3$ is the permutation group over the three spin flavors. The network is then trained by Monte Carlo sampling from the network distribution and updating the weights, using a variant of stochastic reconfiguration~\cite{Chen_2024}, such that the energy is minimized. We further employ a Gutzwiller projection in the sampling procedure, which has been shown to speed up convergence for the $SU(2)$ $t$-$J$-model~\cite{lange2024simulatingtwodimensionaltjmodel}. We choose a Gutzwiller projected Fermi sea as the initial state, which at unit filling represents the exact ground state in the $SU(N\rightarrow\infty)$ limit~\cite{Kawakami_1992}. We compare our results to matrix product state (MPS) calculations~\cite{hubig:_syten_toolk, hubig17:_symmet_protec_tensor_networ} in an open boundary system up to $9\times9$ lattice sites, and find good agreement or even a slight advantage of the NQS method, see SM~\cite{SM}. In the following, we will always focus on fully periodic systems, where MPS with their inherently one-dimensional structure become prohibitively expensive.
\\

\textbf{Magnetic and Polaronic Correlations.} We start by investigating the magnetic properties of the model in Eq.~\ref{eq:H_tJ} at zero doping. Fig.~\ref{fig:1}b shows a correlation map of the $SU(3)$ spins as a function of distance $\vec{d}$ between sites. In analogy to the spin-spin correlations $\langle \hat{S}^z_\mathbf{i}\hat{S}^z_{\mathbf{i+d}}\rangle$ of $SU(2)$ symmetric models, we study $\langle \hat{\lambda}_{3,\mathbf{i}}\hat{\lambda}_{3,\mathbf{i+d}}\rangle$ and $\langle \hat{\lambda}_{8,\mathbf{i}}\hat{\lambda}_{8,\mathbf{i+d}}\rangle$, where $\hat{\lambda}_{8,\mathbf{i}}$, $\hat{\lambda}_{3,\mathbf{i}}$ represent the two diagonal $SU(3)$ generators
\begin{equation}
\hat{\lambda}_3=
    \begin{bmatrix}
        1&0&0\\
        0&-1&0\\
        0&0&0
    \end{bmatrix}, \hspace{15pt} \hat{\lambda}_8=\frac{1}{\sqrt{3}}
    \begin{bmatrix}
        1&0&0\\
        0&1&0\\
        0&0&-2
    \end{bmatrix}
\end{equation}
on site $\mathbf{i}$. We fix the choice of flavor labels such that $\hat{\lambda}_{3,\mathbf{i}}=\hat{n}_{\mathbf{i}.r}-\hat{n}_{\mathbf{i},g}$ and $\hat{\lambda}_{8,\mathbf{i}}=1/\sqrt{3} \ (\hat{n}_{\mathbf{i},r}+\hat{n}_{\mathbf{i},g}-2\hat{n}_{\mathbf{i},b})$. Fig.~\ref{fig:1}b shows the spin-spin correlations

\begin{equation}
\label{eq:SS_corr}
    C_\tau(\mathbf{d}) = \sum_{i}\eta_{\mathbf{i},\mathbf{i+\vec{d}}}\langle \hat{\lambda}_{\tau, \mathbf{i}}\hat{\lambda}_{\tau, \mathbf{i+\vec{d}}}\rangle,
\end{equation}

\noindent for $\tau=3$, where $\eta_{\mathbf{i},\mathbf{i+\vec{d}}}=1/\left[\sigma(\lambda_{\tau, \mathbf{i}})\sigma(\lambda_{\tau, \mathbf{i+\vec{d}}})\right]$, with $\sigma^2(\lambda_{\tau, \mathbf{i}})$ the variance of $\lambda_\tau$ at site $\mathbf{i}$, is a normalization constant that accounts for the presence of holes. Note that due to the combinatorics of the possible three-flavor nearest neighbor states, the range between the maximal values of the correlations is smaller than in the $SU(2)$ analogue. Taking into account the normalization, one finds $-1/2 \leq C_\tau(\mathbf{d}) \leq 1$ for the $SU(3)$ case, see SM~\cite{SM}. Due to the enforced spin permutation symmetry, we know a priori that $\langle \hat{\lambda}_{3,\mathbf{i}}\hat{\lambda}_{3,\mathbf{j}} \rangle=\langle \hat{\lambda}_{8,\mathbf{i}}\hat{\lambda}_{8,\mathbf{j}}\rangle \forall \mathbf{i},\mathbf{j}$. Fig.~\ref{fig:1}c shows a cut through the same correlations with respect to two different reference sites on different sublattices. We see the same 3-SL pattern for both reference sites, confirming the overall 3-SL order of the ground state.

Next we study the evolution of magnetic correlations upon hole-doping $\delta=n_h/L$ for $t/J=3$. Fig.~\ref{fig:2}b shows the magnitude of the nearest neighbor (\textit{nn}) and next-nearest neighbor (\textit{nnn}) correlations for different dopings. The cartoon inset shows the specific distances considered, which correspond to \textit{nn} sites on different sublattices, and \textit{nnn} sites on the same sublattice, showing the expected FM and AFM correlations of the 3-SL order in the low-doping regime. We observe that upon doping the correlations decrease, with the \textit{nnn} correlations going through a sign change at $\delta\approx0.25$. The inset also show a comparison with the analogous \textit{nn} and \textit{nnn} correlations on the $SU(2)$ square lattice from Ref.~\cite{lange2024simulatingtwodimensionaltjmodel}, which we find to be strikingly similar, highlighting the aforementioned analogy between the two models. We find that the sign crossing happens at similar doping values, indicating that the magnetic polaron regime and AFM region of the phase diagram have a similar dependence on doping as in the $SU(2)$ square lattice case. Note that the sign change in the $SU(3)$ case is less pronounced, as expected from the fact that the maximal \textit{nn} AFM correlations are smaller than in the $SU(2)$ case. The origin of the sign change and shape of the correlations can be explained within the geometric string theory developed for the $SU(2)$ Hubbard model, see next section.


To further investigate the effect of hole doping on the AFM order, we also study higher-order magnetic polaron correlations. In analogy to the $SU(2)$ case~\cite{lange2024simulatingtwodimensionaltjmodel,Koepsell2021}, we define the connected correlations

\begin{equation}
    C_\circ^\alpha(\mathbf{d}, \mathbf{d}_h) = \frac{1}{N_dN_{d_h}}\sum_i \tilde{\eta}_{i,d,d_h} \langle \hat{\lambda}_{\alpha,i+d_h}\hat{\lambda}_{\alpha,i+d_h+d}\hat{n}_i^h\rangle_c,
\end{equation}

\noindent describing the spin correlations relative to a doped hole. Here $\langle\hat{n}^h_i\rangle$ is the local hole density, $\langle\rangle_c$ denotes the connected correlation. $N_{d_{(h)}}$ is the number of sites around the hole that are distance $d_{(h)}$ from the site of the hole. The distances considered are shown in the cartoon inset, and we average over all equivalent configurations of sites around the hole. As before we normalize correlations by $\tilde{\eta}_{i,d,d_h}=1/\left[\langle \hat{n}^h_i\rangle\sigma(\hat{\lambda}^3_{i+d_h})\sigma(\hat{\lambda}^3_{i+d_h+d})\right]$. The evolution of the correlations with hole doping is shown in Fig.~\ref{fig:2}c. Upon comparison with the $SU(2)$ square lattice (shown in the inset)~\cite{lange2024architecturesapplicationsreviewneural}, we again find a very similar shape. However, despite the qualitative agreement, we also observe a significant change in the amplitude of the connected correlations in the $SU(3)$ case. This quantitative change can again be understood from the microscopic string picture.
\\

\textbf{Comparison to Geometric String Theory.}
To provide a microscopic explanation of the observed magnetic correlations, we compare our results to the geometric string theory developed in the $SU(2)$ framework~\cite{Grusdt2023, Bohrdt_2022, Grusdt_2019, Homeier_2025, Bohrdt2020}. In this setting, the theory has yielded key insights into doped charge carriers and laid the groundwork for a cuprate pairing mechanism based on a Feshbach resonance. Extending and testing this theory in more general settings is therefore essential. Within the frozen spin approximation (FSA), where slow spin dynamics $(\propto J)$ are neglected relative to faster hole motion $(\propto t)$, doped holes in the low-doping regime generate geometric strings. Upon moving through the AFM background, a hole creates a string of displaced spins, see Fig.~\ref{fig:2}a. This provides a microscopic explanation for the sign change in the nn and nnn correlations: spin displacements along the string effectively flip FM and AFM bonds, mixing correlations and reversing the \textit{nnn} amplitude.

The theory can be extended by analyzing hole motion in a frozen three-flavor AFM (Fig.~\ref{fig:2}a). As in the $SU(2)$ case, we observe a sign change in the two-point correlations for the same reason. However, in $SU(3)$ the shorter correlation range—due to the normalization of the Gell-Mann matrices—reduces the negative mixing contribution, making the sign change less pronounced. The change in the amplitude of the connected $SU(2)$ and $SU(3)$ correlations in Fig.~\ref{fig:2}c is likewise explained by string effects on neighboring bonds. Due to the higher connectivity of the triangular lattice and the additional flavors, the fraction of overall bonds whose configuration changes due to the hole motion from aligned to anti-aligned and vice versa is only $2/6$, while $2/4$ bonds are affected in the $SU(2)$ case. This smaller effect of the hole motion on the overall order of the spin background is reflected by the fact that the connected correlations are significantly smaller in the $SU(3)$ case.
To make the comparison quantitative, we compute correlator expectation values within geometric string theory. Starting from undoped snapshots, we introduce holes with strings of displaced spins. The string-length distribution $p(l)$ is obtained by mapping the single-hole problem to a semi-infinite 1D model (see SM~\cite{SM}). At finite doping, we add the required number of holes, assigning each a string with length drawn from $p(l)$, and evaluate the correlators as in Eq.~\eqref{eq:SS_corr}. The predicted nn correlations agree well with the NQS results (see Fig.~\ref{fig:2}b), while deviations appear in the nnn correlator at intermediate doping.
\\

\textbf{Pair Structure and Binding Energies.}
One of the central open questions in the physics of cuprate superconductors is the microscopic mechanism of pairing.
As a first step to study potential pairing in the $SU(3)$ model, we turn to an exploration of the two-hole structure. Due to the spin imbalance, we do not enforce the spin permutation symmetry in the NQS and work in a sector with fixed magnetization. 

\begin{figure}[t!]
    \centering
    \includegraphics[width=0.48\textwidth]{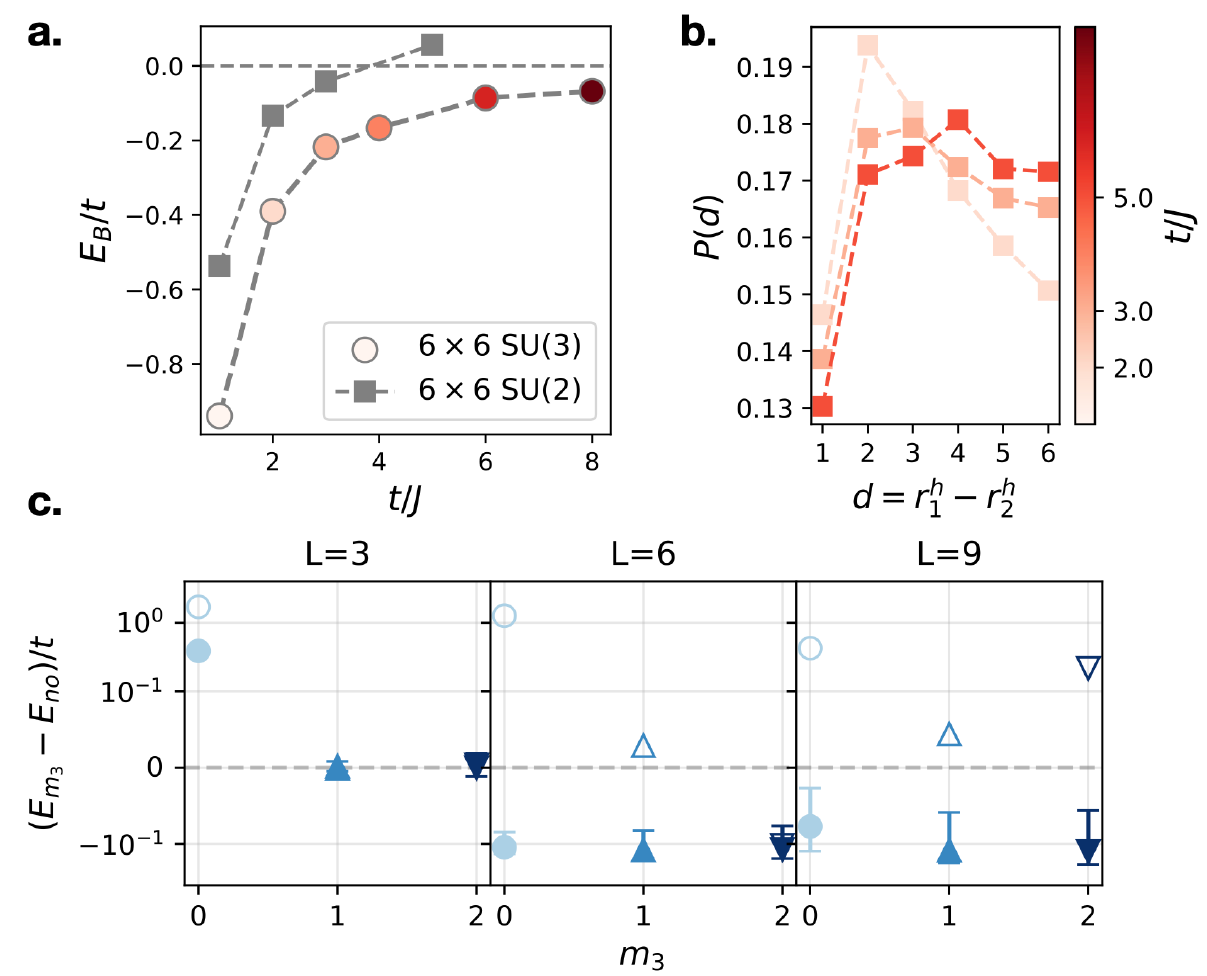}
    \caption{\textbf{a.} Binding energies $E_B$ for different ratios of $t/J$. A negative binding energy indicates a lower energy for the two-hole state. \textbf{b.} Distribution of hole distances $d$ from snapshots of the optimized NQS with two doped holes on a $9\times9$ torus, where the maximal distance between two sites is $d=6$. The total number of snapshots taken into account is $16\times10^3$ for each $t/J$ ratio considered (2,3 and 5, shown as ticks on colorbar). \textbf{c.} Variational energies of the ground state with two doped holes on a $L\times L$ lattice, projected in different rotational symmetry sectors relative to the unsymmetrized optimization, evaluated both from importance sampling from the unsymmetrized ground state (empty markers) and a fully symmetrized optimization (filled markers). Where empty markers are not visible they overlap with the filled markers. Errorbars are smaller than the markers for $3\times 3$.}
    \label{fig:4}
\end{figure}

We start by analyzing the binding energies $E_B = E_{2h}-2E_{1h}$, for different parameters $t/J$. Here, $E_{1h}$ and $E_{2h}$ represent the energies of the single-hole and two-hole states relative to the undoped Néel state. The results are shown in Fig~\ref{fig:4}a. We find a negative binding energy $E_B<0$, the magnitude of which decreases as $t/J$ is increased. This indicates that the doped charge carriers can lower their energy by binding together. Upon comparison, we find the amplitude of the binding energies to be larger than in the $SU(2)$ square lattice case, where our NQS results are further confirmed by large-scale density matrix renormalization group (DMRG) studies, which reveal binding energies on the order of $0.2t$ at $t/J=2$~\cite{blatz2024twodopantorigincompetingstripe} and exact diagonalization studies showing a zero crossing at similar values of $t/J$~\cite{Chernyshev_1998, Leung_2002}.
This large binding energy is especially surprising from a conventional BCS perspective, where the larger connectivity $z$ of the triangular lattice results in a smaller effective electron mass, which in turn indicates a smaller superconducting gap $\Delta$ and pairing energy of the Cooper pairs. However, in the geometric string picture, the increased binding energies can be explained by a geometric spinon-chargon repulsion, which increases with $z$, favoring bound states at higher connectivity~\cite{Grusdt2023, Bohrdt_2022}, see SM~\cite{SM}.

We further investigate the distribution of distances between two doped holes available directly from the Monte Carlo snapshots, as shown in Fig~\ref{fig:4}b. Here we normalize each distance $d$ by the total number of pairs of sites that are distance $d$ apart. We see that, as expected for small $t/J$, the holes tend to be tightly bound, while for larger $t/J$ ratios the average distance increases, suggesting more spatially separated pairs.

We probe the rotational structure of two-hole pairs by analyzing the rotational symmetry sector of the ground state. As evident from Fig.~\ref{fig:1}a, the AFM order reduces the lattice $C_6$ symmetry to $C_3$. We impose this symmetry in the HFDS architecture by averaging over rotations:

\begin{equation}
\label{eq:rotsym}
    \ket{\psi_{m_3}(\sigma)}=\sum_{i=0}^2 e^{i2\pi/3jm_3} \sum_\sigma \tilde{\psi}(\hat{R}^{j}\sigma)\ket{\sigma}
\end{equation}

\noindent where $m_3=0,1,2$ labels irreps, $\hat{R}^j$ rotates $\sigma$ by $120^\circ\times j$, and $\tilde{\psi}$ is the unsymmetrized network output. States with $m_3=1,2$ are degenerate due to time-reversal symmetry.
To identify the symmetry sector of the unsymmetrized state, we train without symmetry constraints and use importance sampling to compute
$E(m_3)=\sum_{s\in\tilde{S}} w(s)E_{\mathrm{loc}}(s)$,
with samples $\tilde{S}$ from $\ket{\tilde{\psi}}$ and weights $w(s)=|\psi_{m_3}(s)|^2/|\tilde{\psi}(s)|^2$. Results (Fig.~\ref{fig:4}c, open symbols) show that the $m_3=0$ projection has consistently higher energy than the non-trivial sectors.

We also perform fully symmetrized optimizations, which require averaging the determinants in Eq.~\eqref{eq:rotsym}, and are therefore computationally more expensive, as well as complicate the optimization for larger systems. While the optimized $m_3=1,2$ states have slightly lower energies, for large systems all three sectors agree within sampling error, see SM~\cite{SM} for details. As a next step, further studies at finite doping will be  necessary to determine the nature of a potential superconducting ground state in this regime. However, the clear result in the $3\times3$ system together with the suggestive data of the projection from the unsymmetrized sector presents a strong indication of a non-s-wave pairing symmetry. 
\\

\textbf{Discussion and Outlook.}
In this work, we have investigated finite doping in the $SU(3)$ $t$-$J$ model on the triangular lattice. Owing to its tripartite structure, the triangular lattice with $SU(3)$-symmetric interactions exhibits a close analogy to the much-studied $SU(2)$ bipartite square lattice.
We introduced a three-flavor extension of Gutzwiller-projected hidden fermion determinant states (G-HFDS), which allows us to explore the full doping range of the $SU(3)$ triangular lattice $t$-$J$ model on system sizes up to $9\times9$ lattice sites with fully periodic boundary conditions.
We analyzed two- and three-point magnetic correlations formulated in terms of the Cartan generators $\lambda_3$ and $\lambda_8$ of the $SU(3)$ group to probe the magnetic response and the nature of doped charge carriers. These observables can be directly probed in cold atom experiments~\cite{Buob_2024, Taie_2022, mongkolkiattichai2025quantumgasmicroscopythreeflavor} and reveal strong analogies with their counterparts on the $SU(2)$ square lattice, supporting our claim of a close correspondence between the two models.

Key phenomena such as the formation of dopant pairs, the resulting spin and charge ordering or the formation of a pseudogap may differ significantly in the $SU(3)$ case. The increased connectivity of the triangular lattice, together with the enlarged local Hilbert space due to the third spin flavor, introduces new complexity into the problem of understanding stripe formation and other forms of charge or spin order.
As a first step toward addressing these questions, we examined binding energies across different parameter regimes and found values significantly above those observed in the $SU(2)$ model, suggesting an enhanced tendency for pairing. This is further supported by our analysis of hole–hole distances in the variational wavefunction, where we observe a clear crossover from tightly bound holes at low $t/J$ to more delocalized behavior at larger $t/J$. We further examined the rotational symmetry of two doped holes, finding a strong indication of a non-s-wave rotational ground state. However, the detailed structure of such bound states and their implications for emergent phases remain open questions. In particular, the nature of two-particle pairing is intrinsically richer in the $SU(3)$ model due to the additional spin flavor~\cite{Rapp_2007} as well as the absence of particle-hole symmetry.

Our results represent an important first step toward understanding finite doping in $SU(N)$ Fermi-Hubbard models, opening a path to exploring doped Mott insulators in higher-symmetry systems beyond the $SU(2)$ paradigm and establishing a universal approach to strongly coupled systems. Our work is directly relevant to current and near-future quantum simulation experiments with ultracold alkaline-earth atoms, where $SU(N)$ symmetries naturally emerge due to the decoupling of nuclear spin from electronic degrees of freedom. In particular, quantum gas microscopy platforms offering site-resolved imaging and control can enable direct measurement of spin and charge correlations in ultracold $SU(N)$ systems~\cite{Taie_2022, gasferrer2026spinresolvedmicroscopy87srsun}.
\\

\textbf{Code availability.} Our implementation of G-HFDS is based on the NetKet package~\cite{netket3:2022}, specifically \url{https://netket.readthedocs.io/en/latest/tutorials/lattice-fermions.html} and can be found on \url{https://github.com/annikaboehler/SU3_GHFDS#}. 

\textbf{Acknowledgements.} 
We thank Hannah Lange, Henning Schlömer, Timothy Harris, Linus Hein, Pit Bermes, Changkai Zhang, Jan von Delft and Kaden Hazzard for useful discussions. A. Böhler acknowledges funding by the Munich Quantum Valley (MQV) doctoral
fellowship program, which is supported by the Bavarian state government with funds from the Hightech Agenda Bayern Plus. This project has received funding from the European Research Council (ERC) under the European Union’s Horizon 2020 research and innovation programm (Grant Agreement no 948141) — ERC Starting Grant SimUcQuam. This project was funded by the Deutsche Forschungsgemeinschaft (DFG, German Research Foundation) under Germany's Excellence Strategy -- EXC-2111 -- 390814868. The authors gratefully acknowledge the scientific support and resources of the AI service infrastructure LRZ AI Systems provided by the Leibniz Supercomputing Centre (LRZ) of the Bavarian Academy of Sciences and Humanities (BAdW), funded by Bayerisches Staatsministerium für Wissenschaft und Kunst (StMWK).

\bibliography{main}

\end{document}